\documentclass[conference]{IEEEtran}
\usepackage{cite}
\usepackage{amsmath,amssymb,amsfonts}
\usepackage{algorithmic}
\usepackage{graphicx}
\usepackage{textcomp}
\usepackage{xcolor}
\usepackage[printonlyused]{acronym}
\usepackage{hyphenat}

\def\BibTeX{{\rm B\kern-.05em{\sc i\kern-.025em b}\kern-.08em
    T\kern-.1667em\lower.7ex\hbox{E}\kern-.125emX}}

\acrodef{LEO}{Low Earth Orbit}
\acrodef{MEO}{Medium Earth Orbit}
\acrodef{GEO}{Geostationary Earth Orbit}
\acrodef{GNSS}{Global Navigation Satellite System}
\acrodef{NTN}{Non-Terrestrial Network}
\acrodef{DSSS}{Direct-Sequence Spread Spectrum}
\acrodef{PNT}{Positioning, Navigation, and Timing}
\acrodef{5G}{Fifth Generation}
\acrodef{6G}{Sixth Generation}
\acrodef{NR}{New Radio}
\acrodef{GPS}{Global Positioning System}
\acrodef{UE}{User Equipment}
\acrodef{RF}{Radio Frequency}
\acrodef{COM}{Communication}
\acrodef{JCAP}{Joint Communication and Positioning}
\acrodef{CRLB}{Cramér–Rao Lower Bound}
\acrodef{OFDM}{Orthogonal Frequency-Division Multiplexing}
\acrodef{ANF}{Adaptive Notch Filter}
\acrodef{RIM}{Robust Interference Mitigation}
\acrodef{SNR}{Signal-to-Noise Ratio}
\acrodef{SIR}{Signal-to-Interference Ratio}
\acrodef{SINR}{Signal-to-Interference-plus-Noise Ratio}
\acrodef{C/A}{Coarse/Acquisition}
\acrodef{E1}{Galileo E1 Signal}
\acrodef{F-PLL}{FLL-Assisted PLL}
\acrodef{PLL}{Phase-Locked Loop}
\acrodef{FLL}{Frequency-Locked Loop}
\acrodef{DLL}{Delay-Locked Loop}
\acrodef{PVT}{Position, Velocity, and Timing}
\acrodef{MC}{Monte Carlo}
\acrodef{CDF}{Cumulative Distribution Function}
\acrodef{AWGN}{Additive White Gaussian Noise}
\acrodef{NAV}{Navigation}
\acrodef{BER}{Bit Error Rate}
\acrodef{TOW}{Time of Week}
\acrodef{SMS2}{Sustainable Multifunctional Satellite Systems}
\acrodef{FNR}{Fonds National de la Recherche}
\acrodef{3GPP}{Third Generation Partnership Project}
\acrodef{CMMB}{China Mobile Multimedia Broadcasting}
\acrodef{SoO}{Signals of Opportunity}
\acrodef{GNSSDO}{GNSS Disciplined Oscillator}
\acrodef{OCXO}{Oven Controlled Crystal Oscillator}
\acrodef{PRN}{Pseudo Random Code}
\acrodef{PDSCH}{Physical Downlink Shared Channel}
\acrodef{SIB19}{System Information Block}
\acrodef{PDCCH}{Physical Downlink Control Channel}
\acrodef{SSB}{Synchronization Signal Block}
\acrodef{RB}{Resource Block}

\begin{document}

\title{When 5G NTN Meets GNSS: Tracking GNSS Signals under Overlaid 5G Waveforms\\

\thanks{(Corresponding author: Idir Edjekouane.)}
}

\author{
\IEEEauthorblockN{Idir Edjekouane, Alejandro González Garrido, Jorge Querol, Symeon Chatzinotas}
\IEEEauthorblockA{\textit{Interdisciplinary Centre for Security, Reliability and Trust (SnT)} \\
\textit{University of Luxembourg}\\
Luxembourg City, Luxembourg \\
\{idir.edjekouane, alejandro.gonzalez, jorge.querol, symeon.chatzinotas\}@uni.lu}
}

\maketitle

\begin{abstract}
\nohyphens{%
Global Navigation Satellite Systems (GNSS) provide the backbone of Positioning, Navigation, and Timing (PNT) but remain vulnerable to interference. Low Earth Orbit (LEO) constellations within Fifth-Generation (5G) Non-Terrestrial Networks (NTN) can enhance resilience by jointly supporting communication and navigation. This paper presents the first quantitative analysis of GNSS tracking and navigation message demodulation under a hybrid waveform where a low-power Direct-Sequence Spread Spectrum (DSSS) component is overlaid on an Orthogonal Frequency-Division Multiplexing (OFDM) 5G downlink. We evaluate a minimally modified GNSS receiver that tracks a legacy Global Positioning System (GPS) L1 Coarse/Acquisition (C/A) overlay aligned with 5G frames while treating the 5G waveform as structured interference. Using Monte Carlo simulations under realistic LEO Doppler dynamics, we analyze the Bit Error Rate (BER) of GPS L1 C/A navigation bits and the subframe decoding probability versus Signal-to-Interference-plus-Noise Ratio (SINR) for multiple Signal-to-Interference Ratios (SIR) and dynamic classes. Results show reliable demodulation across wide SINR ranges for low and medium dynamics, whereas high dynamics impose strict lock limits. These findings confirm the feasibility of Joint Communication and Positioning (JCAP) using a near-legacy GNSS chipset with minimal receiver modifications.
}
\end{abstract}


\section{Introduction}

\ac{GNSS} provide the backbone of \ac{PNT} but remain vulnerable in harsh or interfered environments \cite{b1}. \ac{LEO} constellations offer stronger power, reduced latency, and better geometry, making them promising for resilient navigation.

Various \ac{LEO}-based \ac{PNT} strategies have been investigated \cite{b2}, including dedicated constellations, \ac{SoO}, and hybrid approaches. Among them, \ac{JCAP} reuses communication signals for both connectivity and ranging and has gained attention \cite{b3}. The \ac{3GPP} evolution toward \ac{NTN} in \ac{5G} and beyond enables embedding \ac{PNT} functions in large-scale \ac{LEO} systems.

This paper investigates the integration of legacy \ac{GPS} signals into \ac{5G} \ac{NTN} downlinks. In particular, we analyze \ac{GPS} L1 \ac{C/A} overlay on \ac{OFDM} waveforms, where the communication receiver perceives the overlay as interference, whereas a minimally modified \ac{GNSS} receiver exploits it for positioning. The study focuses on \ac{GNSS}-side performance under the high Doppler dynamics characteristic of \ac{LEO} orbits.

\subsection{Related Work}

Embedding navigation signals into communication waveforms has been explored in various contexts. Early examples include the \ac{CMMB} standard, which integrated \ac{DSSS} signals for positioning alongside \ac{OFDM}-based broadcasting \cite{b4}, \cite{b5}. Although designed for terrestrial use, \ac{CMMB} can be viewed as a precursor to modern \ac{JCAP} concepts.

In the satellite domain, one of the first frameworks combining \ac{5G} \ac{NTN} communication with overlay \ac{PNT} signals was proposed in \cite{b6}, showing that the added navigation component could be treated as noise without compromising throughput. More recent work has analyzed hybrid \ac{DSSS}-\ac{OFDM} waveforms in \ac{LEO} \ac{NTN} scenarios. In \cite{b7}, a joint receiver architecture was presented that exploits an overlaid \ac{DSSS} sequence for synchronization and positioning while maintaining robust \ac{OFDM} communication. In \cite{b8}, multi-beam \ac{LEO} satellites were investigated, highlighting trade-offs between spectral efficiency and positioning accuracy. These studies confirmed the feasibility of \ac{JCAP} and paved the way toward resilient \ac{GNSS}-complementary systems.

Nonetheless, prior work has largely emphasized communication performance, leaving the analysis of \ac{GNSS}-side tracking and navigation message decoding under realistic Doppler dynamics mostly unexplored. Moreover, the high Doppler rates of \ac{LEO} satellites, often reaching hundreds of Hz/s, remain a major challenge for \ac{GNSS} receivers.

\subsection{Contributions}
We study a \ac{JCAP} overlay where a \ac{5G} \ac{NTN} downlink carries a low-power \ac{DSSS} \ac{GNSS} signal. Our contributions are:

\begin{itemize}
    \item \emph{Minimal-modification GNSS receiver.} We demonstrate the feasibility of using a near-legacy \ac{GNSS} receiver, requiring only minor changes, while treating the \ac{5G} waveform as a structured interferer.
    \item \emph{GNSS-centric evaluation under LEO dynamics.} We develop a simulation framework that reports \ac{BER} and subframe success probability versus \ac{SINR} for various \ac{SIR} levels and Doppler dynamic classes.    
    \item \emph{Operational lock limits.} For a LEO at 1200~km altitude, we quantify Doppler-rate thresholds for reliable \ac{NAV} demodulation under low, medium, and high dynamics, thereby delineating practical operating envelopes.
\end{itemize}

Section~II details the system and signal models; Section~III presents simulations and results; Section~IV concludes with future work.

\section{System Description}

\subsection{System Architecture}

\begin{figure}[htbp]
    \centering
    \includegraphics[width=0.9\columnwidth]{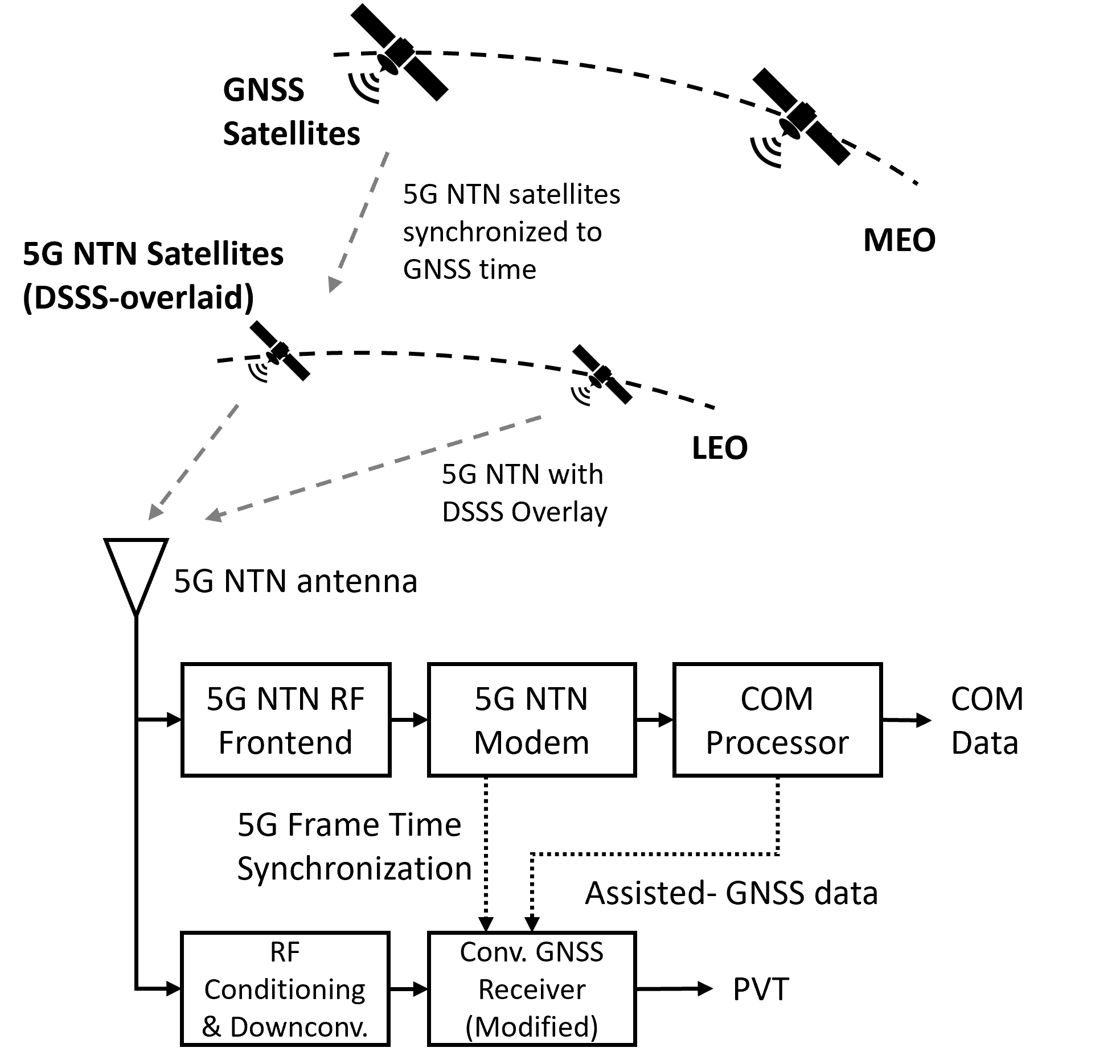}
    \caption{Proposed system architecture for \ac{JCAP} operation in \ac{5G} \ac{NTN}. The received \ac{RF} signal is split between two chains: (\emph{i}) a \ac{5G} modem for broadband communication, and (\emph{ii}) a modified \ac{GNSS} receiver with \ac{RF} conditioning and downconversion. The \ac{5G} downlink includes a low-power \ac{DSSS} overlay synchronized to \ac{GNSS} time, enabling \ac{JCAP} with minimal hardware modification.}
    \label{fig:system_architecture}
\end{figure}

The proposed architecture integrates legacy \ac{GNSS} signals with \ac{5G} \ac{NTN} waveforms, enabling \ac{JCAP} functionalities. The receiver consists of a 5G \ac{NTN} modem operating in parallel with a conventional \ac{GNSS} receiver requiring only minor modifications. Fig.~\ref{fig:system_architecture} illustrates the high-level design.

%A key requirement for any \ac{NTN} constellation offering \ac{PNT} is a precise and stable time reference. Unlike pure communication systems, which tolerate relaxed synchronization, positioning demands nanosecond-level accuracy. Two strategies exist: (\emph{i}) onboard atomic clocks ensuring full autonomy but increasing cost, mass, and power; or (\emph{ii}) synchronization to an external reference derived from \ac{GNSS}. The latter employs a \ac{GNSSDO} combining the long-term stability of \ac{GNSS} with a local high-quality oscillator (typically an \ac{OCXO} or, in higher-grade units, a compact Rubidium oscillator). Such systems can achieve 5--20~ns timing accuracy relative to \ac{GNSS} time and frequency stabilities below $10^{-11}$ over 1~s averaging, providing a cost-effective alternative to onboard atomic clocks at the expense of dependence on external \ac{GNSS} signals~\cite{b9}.

A key requirement for any \ac{NTN} constellation offering \ac{PNT} is a precise and stable time reference. Unlike pure communication systems, which tolerate relaxed synchronization, positioning demands nanosecond-level accuracy. Two strategies exist: (\emph{i}) onboard atomic clocks, enabling full autonomy and independence from external references but increasing cost, mass, and power consumption, and still requiring ground calibration to maintain long-term stability; or (\emph{ii}) synchronization to an external reference derived from \ac{GNSS}. The latter approach yields a constellation dependent on external timing but remains attractive due to its lower cost and maturity. It typically employs a \ac{GNSSDO} combining the long-term stability of \ac{GNSS} with a local high-quality oscillator (typically an \ac{OCXO} or, in higher-grade units, a compact Rubidium oscillator). Such systems can achieve 5--20~ns timing accuracy relative to \ac{GNSS} time, providing a cost-effective alternative to onboard atomic clocks at the expense of dependence on external \ac{GNSS} signals~\cite{b9}.

%On the space segment, \ac{MEO} \ac{GNSS} satellites broadcast navigation signals (e.g., GPS L1 C/A, Galileo E1) serving as time references. In parallel, \ac{LEO} \ac{5G} satellites transmit broadband \ac{OFDM} waveforms with a low-power \ac{DSSS} overlay dedicated to ranging. This overlay can reuse GNSS spreading codes and message formats, enabling backward compatibility with legacy receivers. However, the 16 Keplerian parameters of current messages are insufficient for precise \ac{LEO} orbit description, with errors exceeding 10~m~\cite{b10}. To mitigate this, assistance data can be delivered through the \ac{5G} modem, including precise orbit and clock corrections, almanac or ephemeris updates, similar to assisted-\ac{GNSS} solutions~\cite{b11}. Such assistance data can be conveyed within the \ac{5G} standard framework without requiring an additional application-layer protocol. In \ac{5G}~\ac{NR}, \ac{SIB19} is defined to broadcast \ac{NTN}-related system information, which may include satellite orbit and clock parameters useful for time and frequency assistance. However, in its current form, \ac{SIB19} only supports coarse information, and its format would likely need to be extended or updated to enable the delivery of more precise assistance data.

On the space segment, \ac{MEO} \ac{GNSS} satellites broadcast navigation signals (e.g., GPS L1 C/A, Galileo E1) serving as time references. In parallel, \ac{LEO} \ac{5G} satellites transmit broadband \ac{OFDM} waveforms with a low-power \ac{DSSS} overlay dedicated to ranging. This overlay can reuse \ac{GNSS} spreading codes and message formats, enabling the reuse of existing receiver architectures with minimal modifications. However, the 16 Keplerian parameters of current messages are insufficient for precise \ac{LEO} orbit description, with errors exceeding 10~m~\cite{b10}. To mitigate this, assistance data can be delivered through the \ac{5G} modem, including precise orbit and clock corrections, almanac or ephemeris updates, similar to assisted-\ac{GNSS} solutions~\cite{b11}. Such assistance data can be conveyed within the \ac{5G} standard framework without requiring an additional application-layer protocol. In \ac{5G} \ac{NTN}, \ac{SIB19} is defined to broadcast \ac{NTN}-related system information, which may include satellite orbit and clock parameters useful for time and frequency assistance. However, in its current form, \ac{SIB19} only supports coarse information, and its format would likely need to be extended or updated to enable the delivery of more precise assistance data.

On the ground, the \ac{UE} uses a shared antenna whose received \ac{RF} signal is analogically split between two front-end chains: (\emph{i}) a \ac{5G} modem for broadband communication, and (\emph{ii}) a conventional \ac{GNSS} receiver with minimal adaptations. The split can be implemented using an unequal \ac{RF} power divider, providing most of the input power to the \ac{5G} branch while preserving sufficient power for the \ac{GNSS} chain. The dedicated \ac{RF} conditioning in the \ac{GNSS} branch performs power adaptation (attenuation and limiting) and frequency translation to match the low-power L-band input expected by commercial \ac{GNSS} chips. Although a unified front-end followed by joint sampling would be theoretically possible, the proposed analog split emphasizes practical feasibility by leveraging existing \ac{5G} and \ac{GNSS} chipsets with minimal hardware modification. Quantization and dynamic-range constraints further justify separate signal conditioning to avoid saturation or degraded sensitivity.

We also assume that two consecutive 10~ms \ac{5G} frames (20~ms) are aligned with the \ac{GNSS} navigation bit boundaries. This facilitates bit transition detection, supports longer coherent integration, and improves demodulation, which is critical since bit-edge detection often limits tracking robustness~\cite{b12}.

This cooperative design provides two outputs: broadband communication data from the 5G chain, and \ac{PVT} estimates from the GNSS receiver. By minimizing changes to legacy receivers and reusing existing GNSS signals, the system achieves \ac{JCAP} with reduced complexity while maintaining resilience to interference and spoofing.

\subsection{Signal Model}
The transmitted waveform overlays a legacy \ac{GNSS} signal onto a \ac{5G} \ac{OFDM} downlink, realizing the \ac{JCAP} concept where the communication waveform carries an additional spread-spectrum component. At the transmitter, the relative power between the \ac{GNSS} and \ac{5G} components is controlled by the \ac{SIR}, defined as  
\begin{equation}
    \mathrm{SIR} = \frac{P_{\mathrm{GNSS}}}{P_{\mathrm{5G}}}
\end{equation}
where $P_{\mathrm{GNSS}}$ and $P_{\mathrm{5G}}$ denote the average powers of the respective signals.  

\subsubsection{Transmitter Signal Model}
The continuous-time hybrid signal is expressed as
\begin{equation}
    s(t) = \sqrt{\rho}\,s_{\mathrm{GNSS}}(t) + \sqrt{1-\rho}\,s_{\mathrm{5G}}(t), \qquad 0 \leq \rho \leq 1
    \label{eq:downlink}
\end{equation}
where $s_{\mathrm{GNSS}}(t)$ is a normalized legacy GNSS waveform (e.g., L1 C/A) with $\mathbb{E}\{|s_{\mathrm{GNSS}}(t)|^2\}=1$, and $s_{\mathrm{5G}}(t)$ is a normalized 5G OFDM signal with $\mathbb{E}\{|s_{\mathrm{5G}}(t)|^2\}=1$. The parameter $\rho$ controls the relative power split, with $\rho = P_{\mathrm{GNSS}} / (P_{\mathrm{GNSS}}+P_{\mathrm{5G}})$. Time alignment ensures that the start instant of each \ac{5G} subframe coincides with the start instant of the corresponding \ac{GNSS} subframe.  

\subsubsection{Channel and Received Signal Model}
The transmitted waveform is impaired by LEO-specific Doppler dynamics and additive noise. The instantaneous Doppler frequency is modeled as a second-order polynomial
\begin{equation}
    f_{D}(t) = f_{0} + \dot{f}\,t + \tfrac{1}{2}\ddot{f}\,t^2
    \label{eq:doppler_model}
\end{equation}
where $f_{0}$ is the initial Doppler shift, $\dot{f}$ the Doppler rate, and $\ddot{f}$ the Doppler acceleration.  

Accordingly, the received baseband signal is
\begin{equation}
    r(t) = \Big(\sqrt{\rho}\,s_{\mathrm{GNSS}}(t) + \sqrt{1-\rho}\,s_{\mathrm{5G}}(t)\Big)
            e^{j2\pi f_{D}(t)} + w(t)
\end{equation}
where $w(t)\sim\mathcal{CN}(0,\sigma^2)$ denotes complex \ac{AWGN}. The factor $\rho$ is used exclusively to represent the relative transmit power split between the GNSS overlay and the 5G waveform. Other channel effects such as path loss, fading, or shadowing are neglected here to isolate the Doppler-induced distortion, which is the dominant impairment in \ac{LEO} \ac{NTN} scenarios. These effects can be incorporated in future work by introducing additional channel coefficients or link-budget terms.

The effective reception quality is measured in terms of the \ac{SINR}, defined as
\begin{equation}
    \mathrm{SINR} = \frac{P_{\mathrm{GNSS}}}{P_{\mathrm{5G}} + N_0 B}
\end{equation}
where $N_0$ is the single-sided noise spectral density and $B$ is the receiver bandwidth. 
Unlike \ac{SIR}, which characterizes the transmitted power allocation, \ac{SINR} captures the joint effect of interference and thermal noise at the receiver input.  

This formulation establishes the foundation for the subsequent \ac{MC} simulations later, where the robustness of the tracking and demodulation algorithms is evaluated under varying \ac{SIR}, \ac{SINR}, and Doppler dynamics.

\subsection{GNSS Receiver}
A conventional \ac{GNSS} receiver operates in two stages: acquisition and tracking. Acquisition detects visible satellites and provides coarse code delay and Doppler estimates. Once lock is achieved, the tracking stage refines these estimates, maintains synchronization, and demodulates the \ac{NAV} message. Since this work focuses on \ac{NAV} demodulation, acquisition and \ac{PVT} computation are omitted.

The tracking stage is particularly critical in \ac{LEO} due to the strong Doppler dynamics, which exceed by orders of magnitude those observed in \ac{MEO}. In such conditions, traditional second-order loops are insufficient, as the dynamic stress error quickly exceeds the tracking tolerance. Higher-order loops are therefore mandatory. A third-order \ac{PLL} assisted by a second-order \ac{FLL}, commonly referred to as an \ac{F-PLL}~\cite{b12}, offers a robust compromise: the \ac{FLL} mitigates large frequency errors during high dynamics, while the \ac{PLL} ensures fine carrier phase synchronization. However, third-order \ac{PLL} are inherently sensitive to jerk (third derivative of range), which must be carefully managed through appropriate loop bandwidth settings~\cite{b13,b14}. In parallel, code tracking is maintained by a frequency-aided second-order \ac{DLL}.

A staged tracking strategy is adopted as in~\cite{b13}. It starts with a \emph{pull-in stage}, where loop bandwidths are wide to absorb large errors; then a \emph{coarse-tracking stage} with medium bandwidths; and finally a \emph{fine-tracking stage} with narrow bandwidths for precise \ac{NAV} demodulation. For simplicity, identical loop bandwidths are adopted for the \ac{FLL}, \ac{PLL}, and \ac{DLL} within each stage. The duration of each stage and the associated bandwidth settings (state-machine configuration) are determined empirically through \ac{MC} simulations.

\begin{figure}[htbp]
    \centering
    \includegraphics[width=1\columnwidth]{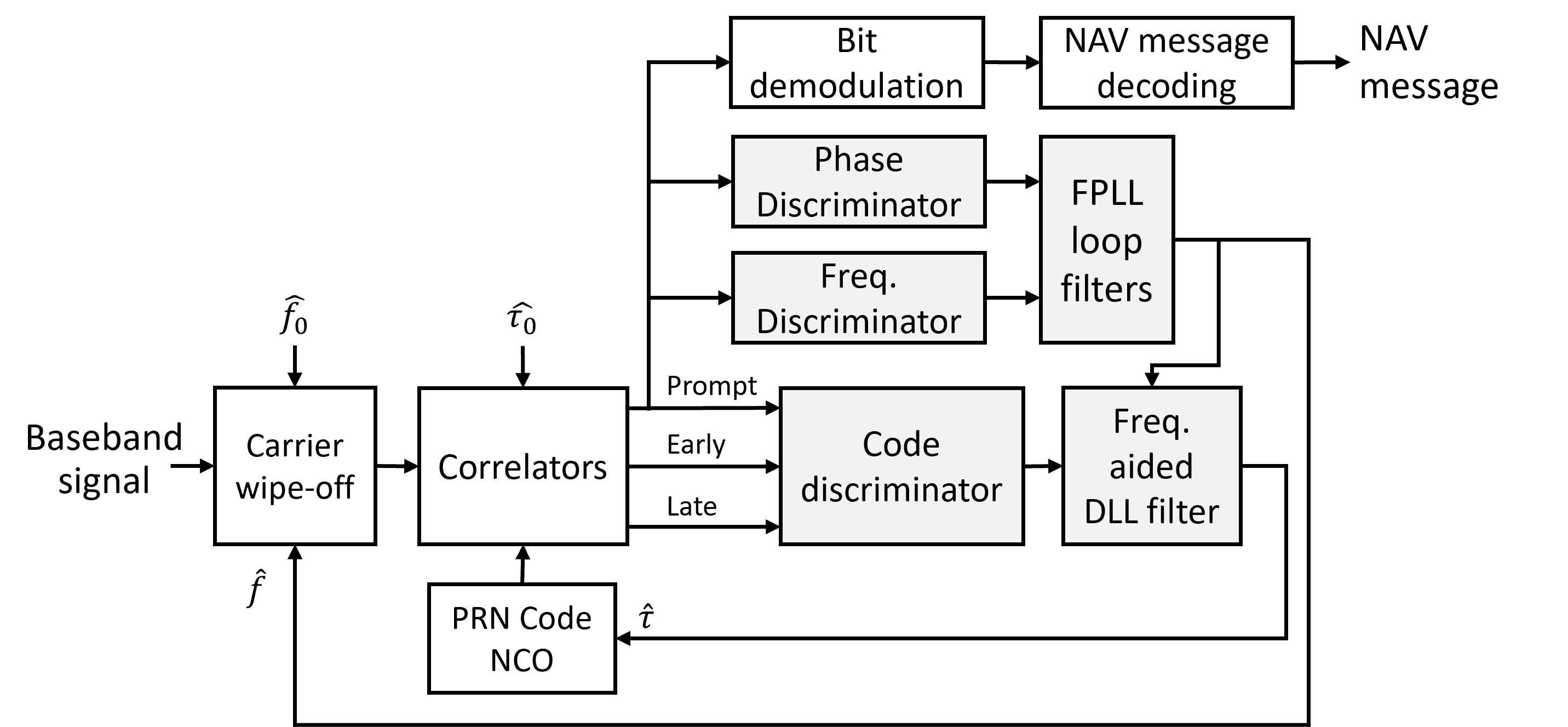}
    \caption{\ac{F-PLL}-based \ac{GNSS} tracking architecture. 
    A third-order \ac{PLL} aided by a second-order \ac{FLL} is used for carrier/phase tracking, 
    while a frequency-assisted second-order \ac{DLL} ensures code tracking. 
    A staged pull-in/coarse/fine strategy enhances robustness under \ac{LEO} Doppler dynamics.}
    \label{fig:gnss_receiver}
\end{figure}

Fig.~\ref{fig:gnss_receiver} shows the adopted architecture. The \ac{F-PLL}, combined with a frequency-assisted \ac{DLL}, achieves reliable \ac{NAV} demodulation under high-dynamic \ac{LEO} conditions with minimal receiver changes.

\section{Simulation Results}
\subsection{Monte Carlo Simulation Description}
\begin{figure}[htbp]
    \centering
    \includegraphics[width=0.7\columnwidth]{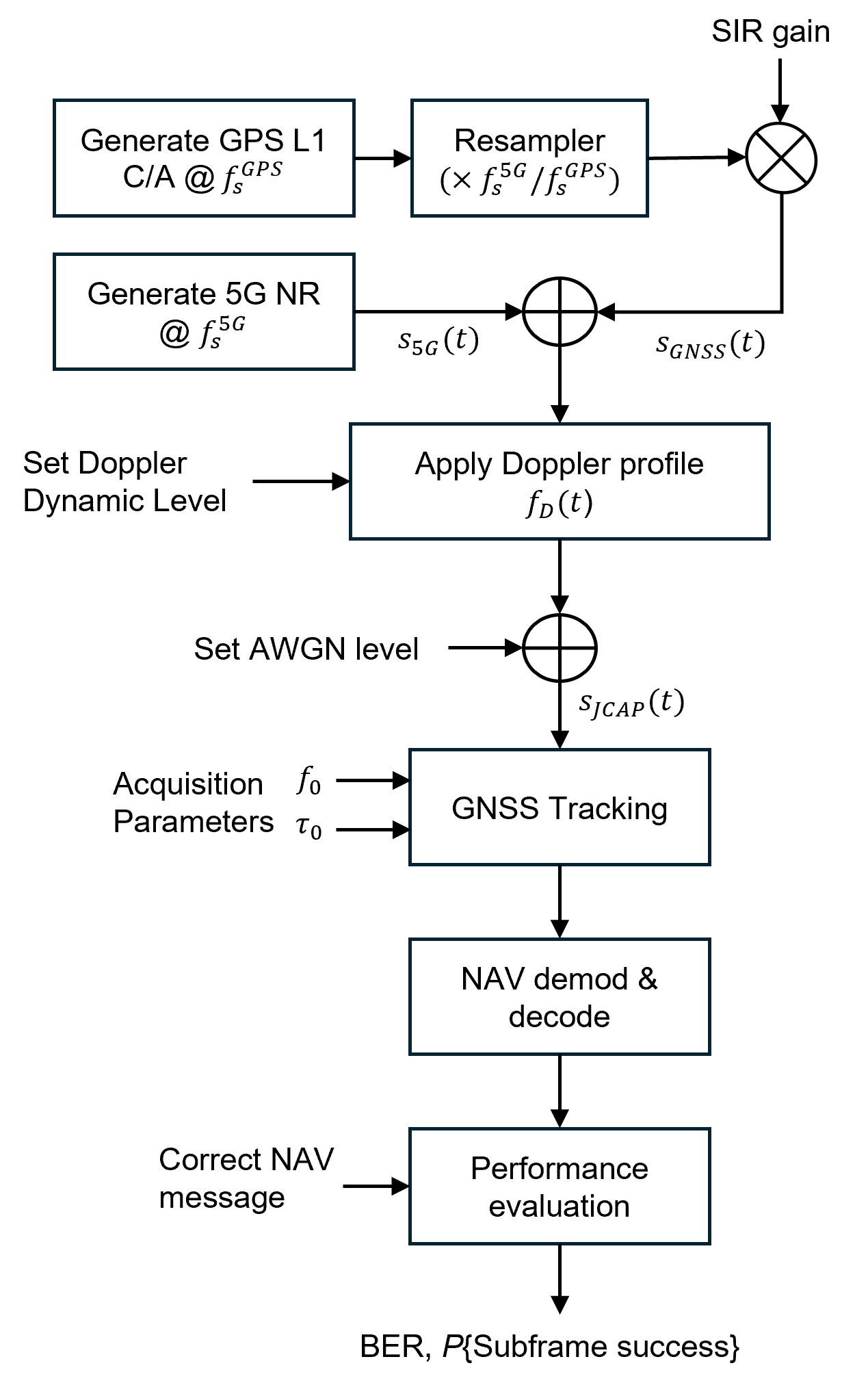}
    \caption{Simulation framework for evaluating the \ac{F-PLL}-based \ac{GNSS} tracking architecture under \ac{5G} interference. The flow includes signal generation, Doppler and noise impairments, tracking, \ac{NAV} message decoding, and performance evaluation.}
    \label{fig:MC_framework}
\end{figure}

\ac{MC} simulations were performed to assess the end-to-end tracking robustness of the \ac{GNSS} receiver under an overlaid \ac{5G} waveform, focusing on the reliable demodulation of the legacy \ac{GPS} \ac{NAV} message. While conventional lock indicators (e.g., phase and frequency lock indicators) can provide useful information about the status of individual loops within the \ac{F-PLL}, they do not fully capture the end-to-end robustness of the tracking process. Instead, monitoring the ability to reliably demodulate and decode the known \ac{NAV} message offers a global and application-oriented metric, directly reflecting the effective performance of the complete tracking chain under realistic \ac{LEO} Doppler dynamics. The overall simulation framework is illustrated in Fig.~\ref{fig:MC_framework}.

In each trial, a legacy \ac{GPS} L1 \ac{C/A} signal was generated using a real \ac{NAV} message obtained from prior \ac{RF} capture. In parallel, a \ac{5G} \ac{NR} waveform corresponding to \emph{Case~A} was generated, including the \ac{SSB}, \ac{PDCCH}, and \ac{PDSCH} channels, where the \ac{PDSCH} carried random data. The total occupied bandwidth was 5~MHz (25~\ac{RB}). This waveform was combined with the \ac{GPS} signal at a predefined \ac{SIR}, ensuring sample-rate alignment via resampling. The resulting \ac{JCAP} signal was then subjected to Doppler dynamics as defined in~\eqref{eq:doppler_model}, with parameters corresponding to low-, medium-, and high-dynamic regimes described in Section~C. Finally, \ac{AWGN} was introduced to achieve the desired \ac{SINR}.

Each simulated sequence had a duration of 10~s. This interval allowed the tracking loops to pass through the pull-in and coarse stages before entering steady fine-tracking operation, ensuring that at least one complete 6~s \ac{GPS} subframe could be decoded. To focus exclusively on tracking and decoding robustness, the acquisition stage was bypassed by providing the correct initial delay $\tau_0$ and Doppler frequency $f_0$.

The tracking stage was implemented using the staged pull-in/coarse/fine configuration described in Section~D. Outputs were then passed to the \ac{NAV} demodulation and decoding module. Two performance metrics were evaluated: (\emph{i}) the \ac{BER} over the full 10~s sequence, and (\emph{ii}) the probability of successfully decoding an entire subframe, which directly reflects the reliability of navigation message reception under the considered interference and Doppler conditions.

The main simulation parameters are summarized in Table~\ref{tab:simu_param}. It is worth noting that, due to the high computational cost of long-duration tracking, the number of \ac{MC} trials was limited to 100.

\begin{table}[htbp]
\centering
\caption{Main Simulation Parameters}
\label{tab:simu_param}
\scriptsize
\begin{tabular}{p{0.42\columnwidth} p{0.42\columnwidth}}
\hline
\textbf{Parameter} & \textbf{Value} \\ \hline
Carrier frequency $f_c$ & 2~GHz \\
5G sampling rate $f_s^{5G}$ & 7.68~MSps \\
GPS sampling rate $f_s^{GNSS}$ & 4.092~MSps \\
SIR levels [dB] & $-10$, $-20$, $-30$ \\
Simulation / Pull-in / Coarse time [s] & 10 / 1.5 / 1.5 \\
Loop bandwidths [Hz] & 18 (pull-in), 7 (coarse), 2 (fine) \\
Loop filter coefficients & Table~8.23 in~\cite{b12} \\
Coherent integration time & 20~ms \\ \hline
\end{tabular}
\end{table}

\subsection{Doppler Dynamic Evaluation}
To characterize Doppler dynamics in \ac{NTN} scenarios, we performed extensive \ac{MC} simulations to extract the statistical distributions of Doppler rate and Doppler acceleration. In line with the reference scenarios defined in 3GPP TR~38.821~\cite{b15}, we considered a \ac{LEO} satellite at an altitude of 1,200~km, corresponding to study case~14 in Table~6.1.1.1-9~\cite{b15}. This case was selected as a representative medium-altitude configuration offering a realistic trade-off between Doppler dynamics and link budget constraints. The configuration operates in the S-band with a carrier frequency of $f_c = 2$~GHz and a handheld terminal.

Satellite orbits were propagated using Keplerian equations over six orbital periods (approximately 600 minutes), with a sampling interval of 15~s. For each trial, a random epoch was selected, and a receiver location was randomly generated, subject to an elevation mask of $[30^\circ, 90^\circ]$. A total of $10^5$ trials were executed to ensure statistically robust results.

It should be noted that \ac{3GPP} has not yet fully specified constellation parameters such as orbital configuration or the number of satellites. To enable a representative simulation framework, we adopt the orbital assumptions proposed in the 3GPP RAN1 contribution~\cite{b16}, which are based on circular polar orbits. These assumptions are in line with current trends in commercial and governmental \ac{LEO} deployments~\cite{b15}.

\begin{figure}[htbp]
    \centering
    \includegraphics[width=0.8\columnwidth]{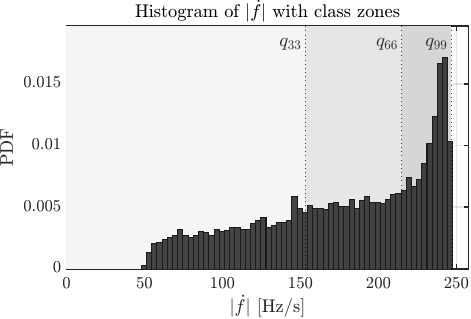}
    \caption{Histogram of Doppler rate $|\dot{f}|$ with quantile thresholds $q_{33}$, $q_{66}$, and $q_{99}$. Shaded regions denote the low-, medium-, and high-dynamic classes adopted for performance evaluation.}
    \label{fig:doppler_rate_hist}
\end{figure}

\begin{figure}[htbp]
    \centering
    \includegraphics[width=0.8\columnwidth]{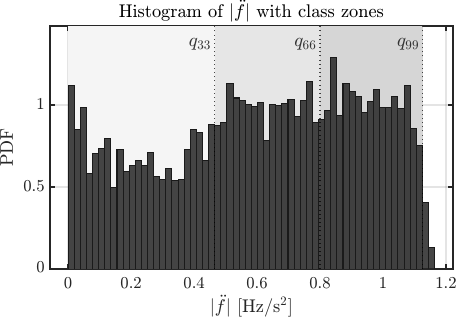}
    \caption{Histogram of Doppler acceleration $|\ddot{f}|$ with quantile thresholds $q_{33}$, $q_{66}$, and $q_{99}$. Shaded regions denote the low-, medium-, and high-dynamic classes adopted for performance evaluation.}
    \label{fig:doppler_acc_hist}
\end{figure}

Fig.s~\ref{fig:doppler_rate_hist} and~\ref{fig:doppler_acc_hist} show the histograms of Doppler rate and Doppler acceleration, respectively. Three Doppler dynamic regions are highlighted for interpretation: (\emph{i}) low dynamics, defined as values below the 33rd percentile ($q_{33}$), (\emph{ii}) medium dynamics, between $q_{33}$ and $q_{66}$, and (\emph{iii}) high dynamics, up to the 99th percentile ($q_{99}$). The representative thresholds reported in Table~\ref{tab:doppler_thresholds} correspond to the maximum values within each category, and are subsequently used when applying the Doppler profile to the generated \ac{JCAP} signal.

\begin{table}[htbp]
    \centering
    \caption{Doppler dynamic thresholds adopted for \ac{JCAP} signal generation at $f_c = 2~\text{GHz}$}
    \label{tab:doppler_thresholds}
    \begin{tabular}{lccc}
        \hline
        \textbf{Metric} & \textbf{Low (L)} & \textbf{Medium (M)} & \textbf{High (H)} \\
        \hline
        Rate [Hz/s]      & 153.01 & 214.90 & 246.59 \\
        Accel [Hz/s$^2$] & 0.47   & 0.80   & 1.13   \\
        \hline
    \end{tabular}
\end{table}

\subsection{Navigation Message Demodulation Evaluation}

This subsection evaluates the ability of the proposed \ac{JCAP} receiver to reliably demodulate the \ac{GPS} \ac{NAV} message in the presence of an overlaid 5G \ac{NTN} waveform. The performance is assessed in terms of the average \ac{BER} and the probability of successful subframe decoding ($P_{\mathrm{sub}}$). A threshold of $P_{\mathrm{sub}} \geq 0.9$ is adopted as a practical benchmark for reliable \ac{NAV} message reception.

\begin{figure*}[htbp]
    \centering
    \includegraphics[width=0.8\textwidth]{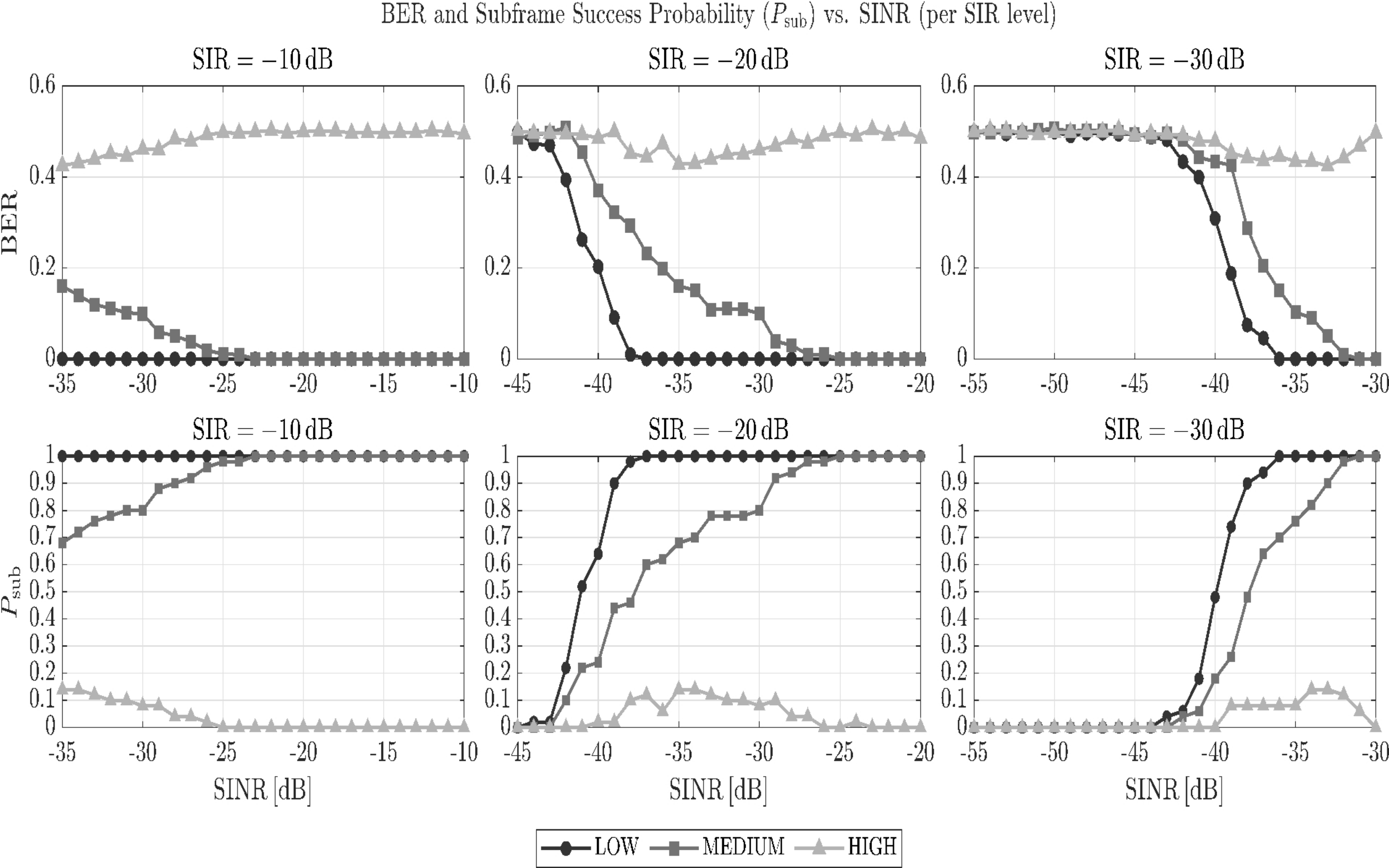}
    \caption{Average \ac{BER} and subframe success probability $P_{\mathrm{sub}}$ versus \ac{SINR} for different \ac{SIR} levels ($-10$, $-20$, and $-30$~dB). The top row shows \ac{BER} performance, while the bottom row illustrates $P_{\mathrm{sub}}$. Results are reported separately for the low-, medium-, and high-dynamic classes defined by Doppler acceleration thresholds $q_{33}$, $q_{66}$, and $q_{99}$.}
    \label{fig:ber_subframe_vs_sinr}
\end{figure*}

Fig.~\ref{fig:ber_subframe_vs_sinr} reports results for three \ac{SIR} levels ($-10$, $-20$, and $-30$~dB), across low-, medium-, and high-dynamic classes defined by the Doppler acceleration thresholds $q_{33}$, $q_{66}$, and $q_{99}$. As expected, the low-dynamic case consistently achieves the best performance, with the $0.9$ threshold attained at significantly lower \ac{SINR} values compared to the other classes. The medium-dynamic case shows moderate degradation but still meets the target at feasible \ac{SINR} levels. In contrast, the high-dynamic case, corresponding to $q_{99}$, exhibits severe degradation, with $P_{\mathrm{sub}}$ rarely exceeding $0.9$. These results confirm that Doppler dynamics are a dominant limiting factor for robust \ac{NAV} message demodulation in \ac{LEO} scenarios.

\begin{figure}[htbp]
    \centering
    \includegraphics[width=0.8\columnwidth]{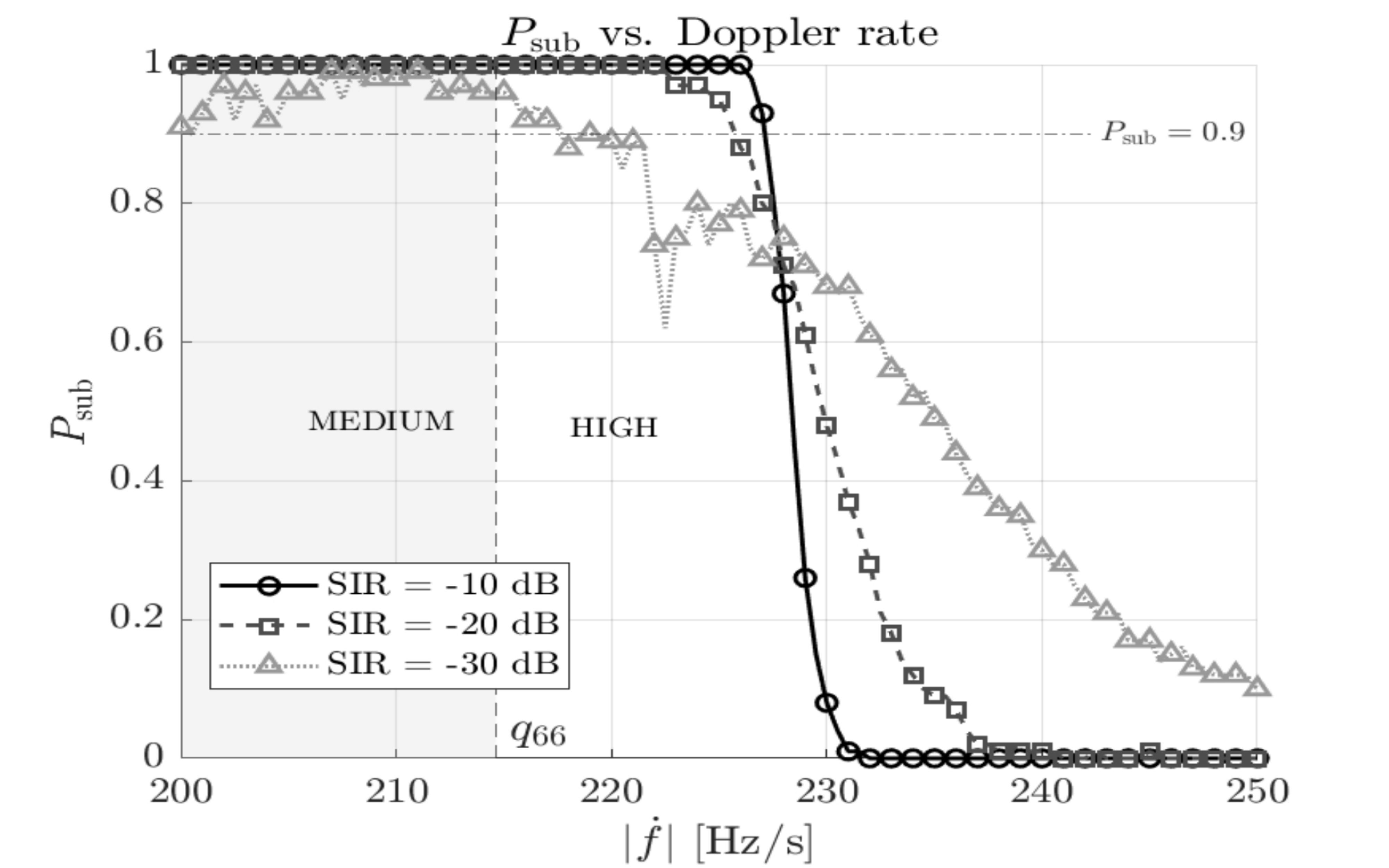}
    \caption{Subframe demodulation success probability $P_{\mathrm{sub}}$ versus Doppler rate for \ac{SIR} levels of $-10$, $-20$, and $-30$~dB. The shaded region denotes the medium-dynamic class ($|\dot{f}| \leq q_{66}=214.63$~Hz/s), while the clear region corresponds to the high-dynamic class ($|\dot{f}| > q_{66}$).}
    \label{fig:psub_doppler}
\end{figure}

To further investigate loop robustness, an additional \ac{MC} experiment was performed focusing on the Doppler rate stress. The Doppler acceleration was fixed to $q_{99}$, and 100 \ac{MC} trials were executed over a Doppler rate range from $-200$ to $-250$~Hz/s, with a resolution of 0.5~Hz/s. Fig.~\ref{fig:psub_doppler} shows the resulting $P_{\mathrm{sub}}$ curves for the three \ac{SIR} levels. The receiver maintains reliable operation ($P_{\mathrm{sub}} > 0.9$) up to approximately 227~Hz/s for the low-dynamic class, 225~Hz/s for the medium class, and 219~Hz/s for the high-dynamic case. Beyond these thresholds, the loops lose lock, leading to abrupt performance degradation.

Overall, these results demonstrate that while low- and medium-dynamic conditions support reliable operation at reasonable \ac{SIR} levels, extreme Doppler dynamics severely constrain \ac{NAV} message demodulation. This limitation motivates the exploration of enhanced tracking strategies capable of maintaining lock under high-dynamic conditions.

\section{Conclusion and Future Work}
We presented a first analysis of \ac{GNSS} tracking and \ac{NAV} demodulation under a hybrid waveform, where a \ac{DSSS} component is overlaid on a 5G \ac{NTN} downlink. Using staged \ac{F-PLL}/\ac{DLL} tracking and 5G–GNSS frame alignment, the receiver achieved high subframe success probability across wide \ac{SINR} ranges under low and medium dynamics, while high dynamics remained the main limiting factor. The derived Doppler-rate lock limits provide concrete guidance for overlay power allocation and aiding strategies.

Future work will compare GPS L1 C/A with Galileo E1, explore additional \ac{LEO} configurations (e.g., 600~km LEO and 300~km V-LEO), and investigate more robust tracking schemes such as vector or deeply-aided architectures, adaptive loop bandwidths, and advanced discriminators. Further attention should also be given to \ac{RF} conditioning techniques required to adapt the received \ac{5G} \ac{NTN} signal to conventional \ac{GNSS} chipsets.

\section*{Acknowledgment}
This work was supported by the project \ac{SMS2} funded by Fonds National de la Recherche (FNR) under Contract C24/IS/18957132/SMS2.

\end{document}